\documentclass[%
 reprint,
superscriptaddress,
 amsmath,amssymb,
 aps,
prl
]{revtex4-2}

\usepackage{graphicx}
\usepackage{dcolumn}
\usepackage{bm}

\usepackage{hyperref}
\hypersetup{pdfstartview={FitH},pdfpagemode={UseNone},
            colorlinks,linkcolor=blue, citecolor=blue, urlcolor=blue,
            bookmarksopen=true, pdfnewwindow=true}
\usepackage[all]{hypcap}

\begin{document}


\title{Broadband bright biphotons from periodically poled triple-resonance metasurface}

\author{Jihua Zhang}
\email{zhangjihua@sslab.org.cn}
\affiliation{Songshan Lake Materials Laboratory, Dongguan 523808, China}
\affiliation{ARC Centre of Excellence for Transformative Meta-Optical Systems (TMOS), Department of Electronic Materials Engineering, Research School of Physics, Australian National University, Canberra, ACT 2600, Australia}

\author{Chaoxin Shi}
\affiliation{Songshan Lake Materials Laboratory, Dongguan 523808, China}
\affiliation{School of Physics, Nanjing University, Nanjing 210093, China}


\author{Jinyong Ma}
\affiliation{ARC Centre of Excellence for Transformative Meta-Optical Systems (TMOS), Department of Electronic Materials Engineering, Research School of Physics, Australian National University, Canberra, ACT 2600, Australia}

\author{Frank Setzpfandt}
\affiliation{Institute of Applied Physics, Abbe Center of Photonics, Friedrich Schiller University Jena, Jena 07745, Germany}

\author{Thomas~Pertsch}
\affiliation{Institute of Applied Physics, Abbe Center of Photonics, Friedrich Schiller University Jena, Jena 07745, Germany}

\author{Chunxiong Bao}
\affiliation{School of Physics, Nanjing University, Nanjing 210093, China}

\author{Jianjun Zhang}
\affiliation{Songshan Lake Materials Laboratory, Dongguan 523808, China}
\affiliation{Institute of Physics, Chinese Academy of Sciences, Beijing 100190, China}

\author{Andrey A. Sukhorukov}
\affiliation{ARC Centre of Excellence for Transformative Meta-Optical Systems (TMOS), Department of Electronic Materials Engineering, Research School of Physics, Australian National University, Canberra, ACT 2600, Australia}


\begin{abstract}
Biphotons from spontaneous parametric down conversion with broad bandwidth are highly wanted in many quantum technologies.
However, achieving broad bandwidth in both frequency and momentum while keeping a high rate remains a challenge for both conventional nonlinear crystals and recently emerging nonlinear metasurfaces.
Here, we address this challenge by introducing a periodically poled triple-resonance metasurface (PPTM) incorporating a nano-grating atop a periodically poled LiNbO$_3$ thin film.
PPTM supports high-Q guided mode resonances at pump, signal, and idler wavelengths meanwhile enabling quasi-phase matching between three guided modes in a broad frequency/momentum range.
The predicted biphoton rate is over 100 MHz/mW with a frequency bandwidth of 165 nm around 1550 nm and a momentum bandwidth of $13^\circ \times 6^\circ$, improving the state-of-the-art by over three orders of magnitude in rate and one order of magnitude in bandwidths. 
This ultrathin broadband bright biphoton source could stimulate system-level miniaturization of various free-space quantum photonic technologies. 
\end{abstract}

\maketitle


Biphotons or photon pairs generated by spontaneous parametric down conversion (SPDC) are 
the engine for a variety of quantum photonic technologies including quantum communication \cite{Gisin:2007-165:NPHOT}, quantum computing \cite{Bennett:2000-247:NAT}, quantum imaging \cite{Shih:2007-1016:ISQE, Moreau:2019-367:NRP}, and quantum sensing \cite{kutas2022QuantumSensing, clark2021SpecialTopic}. In the physical dimension with continuous values such as frequency and momentum, the bandwidth of the biphotons directly relates to the dimensionality of the Hilbert space and the strength of non-classical correlations for a narrow-band pump in SPDC. Therefore, broadband SPDC sources are highly wanted in many quantum technologies \cite{katamadze2022GenerationApplication}.
For example, a broad frequency bandwidth is desired in wavelength-multiplexed quantum communication 
for distributing information to many users \cite{wengerowsky2018} and in quantum optical coherence tomography for a high axial resolution \cite{abouraddy2002QuantumopticalCoherencea, carrasco2004EnhancingAxial, teich2012VariationsTheme}, while broad momentum bandwidth is desired in quantum imaging for a high transverse resolution 
\cite{kviatkovsky2020MicroscopyUndetected, gilabertebasset2019PerspectivesApplicationsa, moreau2018ResolutionLimits}. However, conventional SPDC sources typically relied on nonlinear crystals with a thickness over thousands times of the wavelength \cite{anwar2021EntangledPhotonpair}, limiting the frequency/momentum of the photon pairs to a certain range determined by the phase matching condition \cite{okoth2019MicroscaleGeneration}. 
Meanwhile, the periodic poling technique has been widely used in crystals and waveguides to enable longitudinal phase matching and thus improve the SPDC rate \cite{arie2010PeriodicQuasiperiodic}. However, it is this longitudinal phase matching that fundamentally limits the number of spatial modes \cite{kviatkovsky2020MicroscopyUndetected}.
Therefore, achieving broad bandwidth in both frequency and momentum while keeping 
a high rate
remains a challenge. 

Recently, it was shown that a (sub)wavelength-thick nonlinear film can generate biphotons with broad bandwidth in both frequency and momentum due to the relaxed longitudinal phase matching requirements \cite{okoth2019MicroscaleGeneration, okoth2020IdealizedEinsteinPodolskyRosen}. However, the SPDC rate became much lower than in thick crystals.
A promising approach to amplify the SPDC 
is using optical resonances by either patterning the thin film into nanostructures or integrating nanostructures with the thin film, i.e. making a resonant metasurface \cite{ma2024EngineeringQuantum}. 
In this regard, both local \cite{santiago-cruz2021PhotonPairs, ma2024generationtunablequantumentanglement} and nonlocal \cite{zhang2022SpatiallyEntangled, santiago-cruz2022ResonantMetasurfaces} resonances have been employed. Nevertheless, these resonant metasurfaces all exhibited an intrinsic trade-off between the enhancement factor of the SPDC brightness and the biphotons' frequency/momentum bandwidth.
High enhancement factors require the utilization of resonances with high quality factors, however high-Q resonances typically happen in a narrow frequency range and require extended structures with a nonlocal response, i.e. momentum-dependent resonance frequencies \cite{shastri2023NonlocalFlat, kolkowski2023NonlinearNonlocal}. These features resulted in SPDC enhancement only in a small frequency/momentum range in previous high-Q nonlinear metasurfaces \cite{zhang2022SpatiallyEntangled, santiago-cruz2022ResonantMetasurfaces}.
As a result, the overall SPDC rate which is the integration over a certain frequency/momentum range has been very low no matter using low-Q local or high-Q nonlocal resonances. 
The highest SPDC rate reported so far is less than 10 Hz \cite{zhang2024QuantumMetaphotonics, ma2024generationtunablequantumentanglement}.
A new approach which can release the full potential of meta-optics in realizing ultrathin SPDC sources with broad bandwidth and high rate is yet to be explored.

\begin{figure}
\centering
\includegraphics[width=1\linewidth]{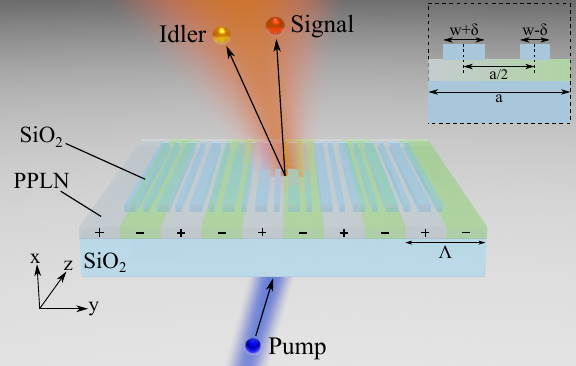}
  \caption{Broadband bright biphoton source from a PPTM consisting of a SiO$_2$ grating (200nm thickness) on a PPLN film (300nm thickness). Inset shows one unit cell of the binary SiO$_2$ grating with a period $a$. 
  The poling period is $\Lambda$. 
  }
  \label{fig:1}
\end{figure}

In this work, we introduce a strategy to generate broadband bright biphotons by a periodically poled triple-resonance metasurface (PPTM) consisting of a binary SiO$_2$ grating atop a periodically poled lithium niobate (PPLN) film (Fig.~\ref{fig:1}). The high rate arises from the simultaneous excitation of high-Q guided mode resonances (GMRs) at all three photon wavelengths and fulfillment of quasi-phase-matching (QPM) between three guided modes. The broad bandwidth is accomplished by realizing the triple-resonance enhanced SPDC in a broad frequency/momentum range, which further improves the rate. 
Under a practical Gaussian pump with a beam waist of $200 \mu m$, the SPDC rate reaches 102 MHz/mW with a frequency bandwidth of 165 nm around 1550 nm and a momentum bandwidth of $13^\circ \times 6^\circ$ relative to the pump direction. The rate and frequency/momentum bandwidths are more than 1000 and 10 times larger than previous metasurface SPDC sources, respectively (Supplementary Table S1). The frequency bandwidth is also comparable to those of previous SPDC sources from van der Waals films without resonances (Supplementary Table S2).
The related number of spatial modes is much larger than that of SPDC sources using nonlinear crystals and waveguides, establishing PPTM as an ideal platform for free-space applications. 
Lastly, the biphotons show a strong momentum anti-correlation within the bandwidth space. 
%
These properties make PPTM a miniaturized and versatile source for numerous quantum photonic technologies such as quantum communications, quantum imaging, and sensing.

%

The role of the linear SiO$_2$ grating is to realize efficient coupling between free-space light and guided waveguide modes in the LN film by exciting GMR, which can enhance the guided field in the LN film from a free-space field or inversely enhance the far-field emission of photons from a waveguide mode. This role is similar to our previous work for double-resonance enhanced SPDC \cite{zhang2022SpatiallyEntangled, ma2023PolarizationEngineering}. The novelty in this work lies in two aspects. 
On one hand, here we not only excite GMRs at the biphoton wavelengths (i.e. signal and idler), 
but also at the pump wavelength, 
realizing the so-called triple-resonance SPDC. 
On the other hand, we employ a binary grating design to simultaneously optimize the input coupling of the GMR at the pump wavelength and the output coupling of the GMRs at the biphoton wavelengths during the SPDC. Specifically, there are two stripes in one unit cell of the grating and a width perturbation $\delta$ is introduced between two stripes, as shown in the inset of Fig.~\ref{fig:1}. This $\delta$ tunes the relative magnitude of the first and second Fourier orders of the grating, and in turn it controls the excitation efficiency and quality factor of the GMRs at the biphoton and pump wavelengths. 
For example, for the extreme case of $\delta=0$, the period of the grating becomes $a/2$ and the original first grating order vanishes. In this case, one cannot excite the GMRs at the biphoton wavelengths, or you can say the GMRs belong to the bound state in the continuum (BIC) mode with an infinite quality factor. With a small perturbation, one can excite the quasi-BIC mode with ultra-high quality factor \cite{huang2023UltrahighQGuided}. 
Therefore, to attain efficient triple-resonance SPDC in this work, a careful design of the width perturbation is essential. 

For a pump light with a specific frequency coming from a specified direction, the linear grating couples it to a transversely propagating waveguide mode with strong field enhancement through the GMR. The role of the nonlinear PPLN grating is to satisfy the QPM condition between this pump waveguide mode and two other waveguide modes at the signal/idler wavelengths in the SPDC process, leading to efficient generation of photon pairs propagating in the waveguide. This is a widely used approach in LN integrated photonics \cite{zhao2020HighQualitya, javid2021UltrabroadbandEntangled}. Finally, the same linear grating couples the photon pairs in the waveguide modes to free-space plane waves with momentum correlations. With the combined effects of linear and nonlinear gratings, we are able to amplify the SPDC in a broad band in both the frequency and momentum domains by leveraging the special dispersion of the GMRs in the metasurface, as will be detailed later. 

Based on the general concept above, the specific dimension of the gratings depends on the desired photon-pair wavelength and emission direction. To generate photon pairs in the telecommunication wavelength band near 1550 nm, we design the dimensions to be $a=810~nm$, $w=0.25a=202.5nm$, and $\delta=75~nm$.
The condition of exciting the GMRs is:
\begin{equation}
    \left(k_y + m \frac{2 \pi}{a}\right) \hat{y} + k_z \hat{z} = \boldsymbol{\beta}_g \,.
\end{equation}
Here $k_{y,z}$ are the transverse wavenumbers along the \textit{y} and \textit{z} directions in the far field. $m=\pm 1, \pm 2 , \cdots$ is the grating order where the signs $\pm$ indicate waveguide modes propagating along the $\pm y$ directions. 
$ \boldsymbol{\beta}_g$ is the wavevector of the guided waveguide mode.
In this work, we only consider the TE modes which have a dominant electric component along the \textit{z} direction such that we can take advantage of the highest nonlinear tensor component $d_{33}$. In addition, all the modes are the fundamental TE modes to ensure a high spatial overlap of the modes in the \textit{x}-\textit{z} plane. 

\begin{figure}
\centering
\includegraphics[width=1\linewidth]{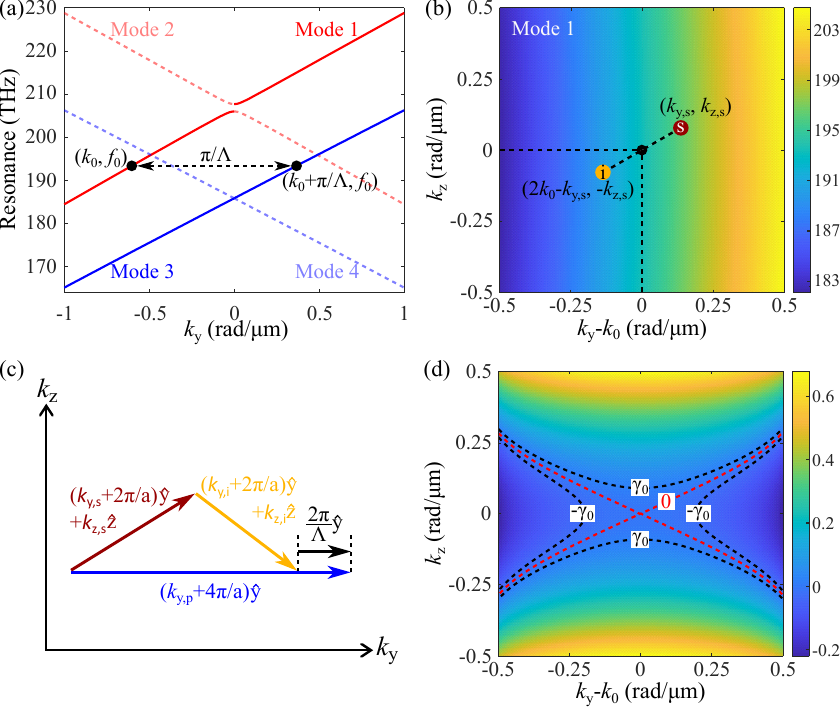}
  \caption{
  (a) Resonance frequency of the first- (Mode 1 \& 2) and second-grating-order (Mode 3 \& 4) GMRs as a function of the transverse wavenumber $k_y$ when $k_z=0$ for the metasurface with $a=810~nm$, $w=0.25a$, and $\delta=75~nm$. For a clear description of the QPM, a transformed plot of Modes 3 \& 4 is applied where both the horizontal and vertical axes are halved. 
  The two black dots mark out the coordinates of two GMRs satisfying the QPM condition for the degenerate SFG/SPDC. 
  (b) Resonance frequency of Mode 1 in the 2D transverse wavenumber space near the degenerate point. Non-degenerate triple-resonance SFG/SPDC is supported in the extended momentum space, where the QPM between the wavevectors of three waveguide modes is shown in (c).
  (d) Plot of $f_1(k_y,k_z)+f_1(2k_0-k_y,-k_z)-2f_0$ near the degenerate point. Dashed red lines represent the contour with zero values and thus perfect QPM. Dashed black lines represent the contours corresponding to the linewidth of the resonances, within which efficient SFG/SPDC are still expected.
  }
  \label{fig:2}
\end{figure}

We define the two first-grating-order modes as \textit{Mode 1} for $m=+1$ and \textit{Mode 2} for $m=-1$. 
These two modes have been studied in our previous work to amplify the SPDC by setting the signal and idler photons on them, realizing the double-resonance SPDC \cite{zhang2022SpatiallyEntangled}. 
They have a linear angular dispersion along the \textit{y} direction (red lines in Fig.~\ref{fig:2}a) 
and a near-flat angular dispersion along the \textit{z} direction (Fig.~\ref{fig:2}b). Two modes are symmetric to each other along the $k_y=0$ plane. In this work, we further investigate the second-grating-order modes, as defined by \textit{Mode 3} for $m=+2$ and \textit{Mode 4} for $m=-2$. Their angular dispersion along the \textit{y} direction is shown in Fig.~\ref{fig:2}a by blue lines, where the values of the horizontal and vertical axes are halved. Modes 3 \& 4 have similar angular dispersion and symmetry properties to Modes 1 \& 2. Therefore, we will only consider the modes propagating along the $+y$ directions in the future. All the results obtained in this work are also applicable when flipping the $k_y$ direction. 

We will target to design a PPTM supporting triple-resonance SPDC where the signal/idler photons are on Mode 1 and the pump photon is on Mode 3. 
In the case of $k_z=0$ in Fig.~\ref{fig:2}a, this means finding two points on the solid red line ($k_s,f_s$) \& ($k_i,f_i$) and one point on the solid blue line ($k_p,f_p$) such that they satisfy the phase matching ($k_s+k_i=k_p$) and energy matching ($f_s+f_i=f_p$) conditions in the SPDC or the inverse sum frequency generation (SFG). As there is no crossing between Mode 1 and the transformed plot of Mode 3, these conditions cannot be fulfilled directly. To do that, we introduce a nonlinear grating which can provide a transverse wavevector $(2\pi/\Lambda)\hat{y}$ to compensate the phase mismatch. In fact, for any point on the solid red line ($k_0,f_0$), we can find a corresponding poling period such that the point ($k_0+\pi/\Lambda,f_0$) is on the solid blue line. The QPM condition for the degenerate SFG/SPDC between these two points is fulfilled.
In this work, for the feasibility of simulating the SFG process using periodic boundary conditions, we choose a point where the needed poling period is an integer multiple of the linear grating period. Specifically, we choose $k_0=-0.6~rad/\mu m$, $f_0 = 193.4~THz$ and the needed poling period is $\Lambda=4a=3.24~\mu m$. Note that here the waveguide modes corresponding to signal/idler photons both propagate along the $+y$ direction, making the current design fundamentally different from the previous studies where two modes propagate along opposite directions \cite{zhang2022SpatiallyEntangled}. 

Most interestingly, as Mode 1 has a weak dispersion along $k_z$ and a linear dispersion along $k_y$, as shown in Fig.~\ref{fig:2}b, the triple-resonance SFG/SPDC with QPM can be realized in a broad range near the degenerate point for the same pump. The related QPM between the wavevectors of the three waveguide modes is shown in Fig.~\ref{fig:2}c. To be more specific, Figure \ref{fig:2}d plots the values of $f_1(k_y,k_z)+f_1(2k_0-k_y,-k_z)-2f_0$ for Mode 1 near the degenerate point. The zero value indicates the region with energy matching, which is marked by the dashed red lines. On top of that, due to a finite quality factor of GMR for Mode 1, efficient triple-resonance SFG/SPDC can be realized in an extended region near the dashed red lines. The area of this region is approximately determined by the linewidth of the GMR, which 
is $\gamma=f/2Q$ where $Q$ is the quality factor of the resonance. Note that in the studied region in Fig.~\ref{fig:2}b, the quality factors of Mode 1 are similar (Fig.~S1), so do the linewidths. The dashed black lines in Fig.~\ref{fig:2}d mark the value of $\gamma_0$ at the degenerate point. Therefore, we expect to see a triple-resonance enhanced and homogeneous SPDC emission in the wide momentum space within these lines. Furthermore, the frequency of photon pairs at each momentum point is close to the resonance frequency of that point in Fig.~\ref{fig:2}b, making the photon-pair source also broadband in the frequency domain.

\begin{figure}
\centering
\includegraphics[width=1\linewidth]{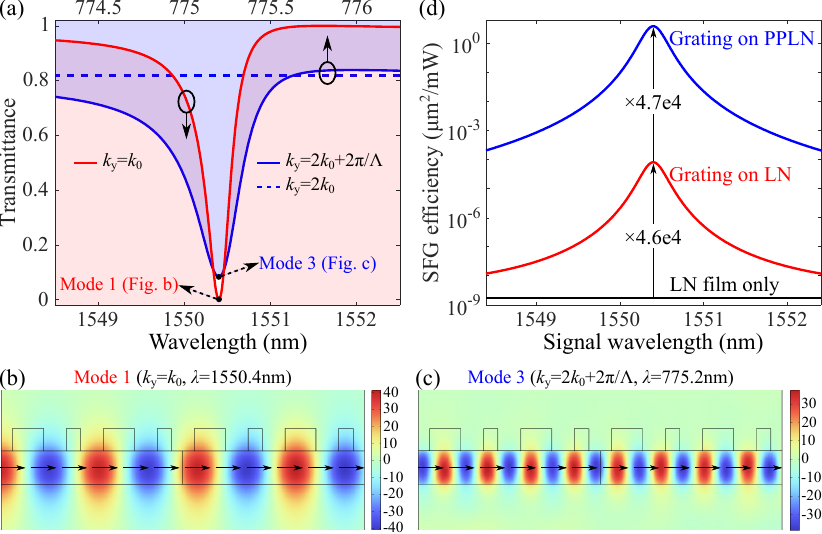}
  \caption{
  (a) Transmission spectra of the metasurface at $k_y=k_0$ near $\lambda_0=1550.4nm$ (solid red line), at $k_y=2k_0+2\pi/\Lambda$ (solid blue line) and $k_y=2k_0$ (dashed blue line) near $\lambda_0/2$.
  (b, c) Enhanced $E_z$ field distributions of Mode 1 \& 3 at their transmission dips. 
  Black arrows represent the direction of the Poynting vectors.
  (d) SFG efficiencies as a function of the signal wavelength for the grating on PPLN with triple resonances, grating on un-poled LN with double resonances, and LN film without resonance. In all simulations, the transverse wave-vectors of the signal and idler light are fixed at $(k_0,0)$ and the SFG frequency is fixed at $2f_0$.}
  \label{fig:3}
\end{figure}

Next, the SPDC is numerically studied by the quantum classical correspondence theory, which relates the SPDC at each momentum point to the inverse SFG process \cite{poddubny2016GenerationPhotonPlasmon, parry2021EnhancedGenerationb,zhang2022PhotonPair}.  To begin with, let's focus on the degenerate point where $k_y=k_0$ and $k_z=0$. The solid red line in Fig.~\ref{fig:3}a is the simulated transmission spectrum for this direction. A sharp transmission dip at the wavelength 1550.4~nm indicates a GMR with a high quality factor of $\simeq 4300$
Correspondingly, the incident field is enhanced by over 40 times inside the LN film at this resonance wavelength (Fig.~\ref{fig:3}b). 
The excited waveguide mode propagates along the $+y$ direction as we expected. Such a high-Q GMR can boost the SFG efficiency. Let's first investigate the SFG without the nonlinear poling grating. To do that, we fix the SFG output with a frequency $2f_0$ and a far-field transverse momentum $(2k_0,0)$. The dashed blue line in Fig.~\ref{fig:3}a tells that there is no resonance at the SFG frequency. We simulate the SFG efficiency at different signal wavelengths, as shown in Fig.~\ref{fig:3}d by the solid red line. One can see that the SFG is boosted by $4.6 \times 10^{4}$ times over the pure LN film (solid black line) at the resonance. 

Then we consider the PPTM with PPLN film. In this case, the SFG output has a shifted momentum $(2k_0+2\pi / \Lambda,0)$. The simulated transmission spectrum at this momentum is shown by the solid blue line in Fig.~\ref{fig:3}a, which indicates a GMR at 775.2 nm with a quality factor of $\simeq 2800$
and a field enhancement of 37 times inside the LN film (Fig.~\ref{fig:3}c). 
The excited waveguide mode also propagates along the $+y$ direction as we expected. With this high-Q GMR at the SFG frequency, the SFG efficiency is further enhanced by $4.7 \times 10^4$ times and reaches 3.76 $\mu m^2/mW$ at the triple-resonance point, as shown by the solid blue line in Fig.~\ref{fig:3}d. Therefore, the inverse SPDC in the same direction will be amplified by the same amount according to the quantum classical correspondence theory. 
We note that it is the combined effect of enhanced generation of waveguide mode by QPM and the enhanced far-field coupling by GMR that leads to the more than four orders of magnitude enhancement of the SFG efficiency at the triple-resonance case over the double-resonance case (see Supplementary Sec. 1).

\begin{figure}
\centering
\includegraphics[width=1\linewidth]{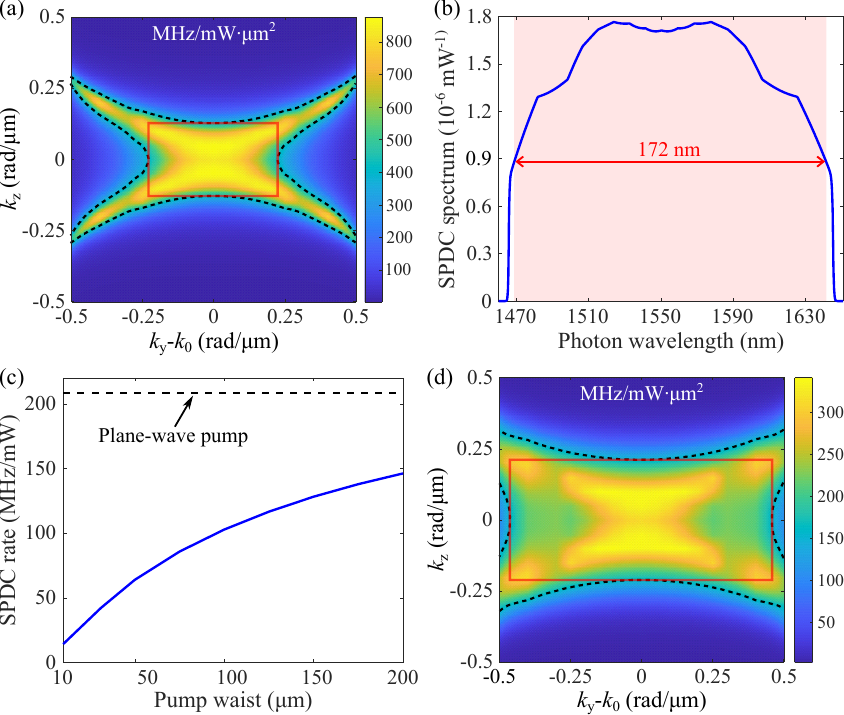}
  \caption{
  (a) SPDC emission pattern and (b) spectrum for a plane-wave pump with frequency $2f_0$ and transverse wave-vector $(2k_0+2\pi/\Lambda, 0)$. 
  (c) SPDC rate for a Gaussian pump with different beam waists. 
  (d) SPDC emission pattern for a Gaussian pump with a beam waist of 200 $\mu m$.
  In (a) and (d), the dashed black lines are contours where the SPDC efficiency is half of the one at the center point and the red rectangles mark the flat emission region defined in this work.}
  \label{fig:4}
\end{figure}

By using the coupled mode theory method \cite{zhang2022PhotonPair}, we are able to efficiently simulate the SFG and thus SPDC in all frequency and momentum combinations for the fixed pump condition above. The calculated momentum and frequency spectra of the SPDC are shown in Figs.~\ref{fig:4}a and~\ref{fig:4}b, respectively. The momentum distribution shows a wide and continuous emission range near the degenerate point. The dashed black lines represent the region where the SPDC efficiency is half of the one at the degenerate point. This region resembles very much the one defined by the linewidth of the GMR in Fig.~\ref{fig:2}d. We define a rectangular flat emission region where the SPDC efficiency is always more than half of its maximum value, which is shown by the red rectangle in Fig.~\ref{fig:4}a. It spans a big momentum space 
near the degenerate point, corresponding to an angular range of $6.4^\circ \times 3.6^\circ$ at the wavelength of 1550~nm. 
The frequency bandwidth in Fig.~\ref{fig:4}b is also broad, having a full width at half maximum of 172 nm. 
The total SPDC rate 
is up to 208 MHz/mW. Within the flat emission region, the SPDC rate is still on the level of 86 MHz/mW. The frequency of photon pairs will be related to the resonance frequencies of the GMRs in this region, resulting in a bandwidth of 81.3 nm (Fig.~S3).

In practice, the pump typically has a Gaussian beam shape instead of the plane-wave shape we have considered above. A Gaussian beam can be expanded by many plane waves from different directions with weighted amplitudes.  
We check that the high performance of the proposed PPTM is robust to the incident direction of a plane-wave pump, where the SPDC rate drops by less than half in an angular range of $(-0.1^\circ,0.07^\circ)$ along $k_y$ and $(-5^\circ,5^\circ)$ along $k_z$ near the designed point (Fig. S5).
This robustness makes it highly practical to use a Gaussian pump. Figure \ref{fig:4}c shows the SPDC rate for a Gaussian pump with different beam waists. 
Although the rate reduces from the plane-wave pump, 
it is still as high as 14.5 MHz/mW under a tightly focused pump beam with a beam waist of 10 $\mu m$. For a practical beam waist of 200 $\mu m$, the rate 
is 146 MHz/mW. Interestingly, the flat emission region spans an even larger momentum space 
relating to an angular range of $13^\circ \times 6^\circ$ at 1550 nm (Fig.~\ref{fig:4}d). The frequency bandwidth is 165~nm (Fig.~S4). The rate within the flat emission region is 102 MHz/mW. In this regard, a Gaussian pump beam is superior to the plane-wave pump.  

\begin{figure}
\centering
\includegraphics[width=1\linewidth]{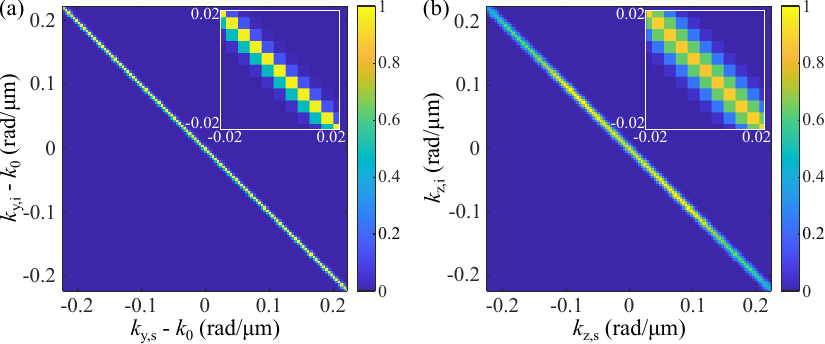}
  \caption{
  Normalized joint momentum intensity of the photon pairs along $k_y$ and $k_z$ for a Gaussian pump with a beam waist of 200 $\mu m$. Insets show the enlarged view near the center.
   }
  \label{fig:5}
\end{figure}

For a Gaussian pump, the momentum correlations between the signal and idler photons typically become weaker \cite{grice2011SpatialEntanglement}. 
To check this, we calculate the joint momentum intensity of the photon pairs along $k_y$ and $k_z$ for a Gaussian pump with a waist of 200 $\mu m$, as shown in Fig.~\ref{fig:5}. 
Even under a Gaussian pump, the photon pairs from the PPTM possess strong momentum anti-correlation along both directions in a continuous and wide range. Note that they are better than the ones from a nonlinear film or crystal due to the resonance-induced momentum selectivity, particularly along the $k_y$ direction due to a stronger angular dispersion (see details in Supplementary Sec.~3). This makes it an ideal photon-pair source for quantum photonic technologies based on momentum anti-correlation or entanglement, e.g. quantum imaging. 

As mentioned before, the QPM for the triple-resonance can be satisfied for any point on the Mode 1 dispersion line in Fig.~\ref{fig:2}a by choosing the appropriate poling period and pump condition. This allows to tune the center frequency and emission direction of the photon pairs in a wide and continuous range. The SPDC rate can be further enhanced by using a smaller perturbation. Experimental realization of the proposed PPTM and its high-performance SPDC is also feasible (see details in Supplementary Secs.~4 \& 5).

To conclude, we have proposed a PPTM to realize an ultrathin, broadband, and bright biphoton source. 
The hybrid two-grating approach breaks the trade-off of previous resonant metasurface SPDC sources between the enhancement factor and bandwidth. 
Furthermore, the generated photon pairs have strong non-classical correlations in the frequency and momentum domains. This photon-pair source could facilitate the miniaturization and integration of numerous quantum photonic technologies including quantum communications, quantum imaging, quantum sensing, and quantum spectroscopy. 


\section{Acknowledgments}
The authors acknowledge support from Guangdong Provincial Quantum Science Strategic Initiative (GDZX2303004, GDZX2200001), Songshan Lake Materials Laboratory (XMYS20230020, 24D1281G111), Innovation Program for Quantum Science and Technology (2021ZD0302300) (J.Z., C.S., J.Z.), 
the Australian Research Council (CE200100010) (J.Z., J.M., A.A.S.), the Deutsche Forschungsgemeinschaft (DFG, German Research Foundation) through the International Research Training Group (IRTG) 2675 “Meta-ACTIVE”, project number 437527638 (T.P., F.S., A.A.S.) and the Collaborative Research Center (CRC) 1375 “NOA” (T.P., F.S.).

\nocite{*}

\bibliography{Bibfile}

\end{document}